\documentclass[12pt,draftclsnofoot,onecolumn]{IEEEtran}
\usepackage{graphicx,cite,epsfig,amssymb,amsmath,subfigure,url,stfloats,latexsym}
\usepackage{multirow,array}
\usepackage{graphicx,cite,amsmath,amssymb,hhline,booktabs,stfloats,bm}
\usepackage{verbatim}
\usepackage[utf8]{inputenc}
\usepackage{subfigure}
\usepackage{algorithm,algorithmic}

\begin{document}
\markboth{IEEE Wireless Communications, Vol. XX, No. YY, Month 2016}
{Wang, Yang, Chen \& Zhang: Computation Diversity \ldots}

\title{\mbox{}\vspace{1.5cm}\\
\textsc{Computation Diversity in Emerging Networking Paradigms} \vspace{1.5cm}}
\author{Kezhi Wang, \textit{IEEE Member}, Kun Yang, \textit{IEEE Senior Member}, $\mbox{Hsiao-Hwa~Chen}${$^{^\dagger}$}, \textit{IEEE Fellow}, and Lianming Zhang
\thanks{Kezhi Wang (e-mails: {\tt kezhi.wang@essex.ac.uk}) and Kun Yang (e-mails: {\tt kunyang@essex.ac.uk}) are with the School of Computer Sciences and Electrical Engineering, University of Essex, CO3 4HG, Colchester, UK. Hsiao-Hwa Chen (email: {\tt hshwchen@mail.ncku.edu.tw}) is with the Department of Engineering Science, National Cheng Kung University, Taiwan. Lianming Zhang (e-mail: {\tt zlm@hunnu.edu.cn}) is with the College of Physics and Information Science, Hunan Normal University, China.}
\thanks{The paper was submitted on May 1, 2016}\\
\vspace{1.5cm}
}

\begin{comment}
\thanks{The paper was submitted on May 1, 2016, and revised on \today.}
\underline{{$^{^\dagger}$}Corresponding Author's Address:}\\
$\mbox{Hsiao-Hwa~Chen}$\\
Department of Engineering Science\\
National Cheng Kung University\\
1 Da-Hsueh Road, Tainan City, 70101 Taiwan\\
Tel: +886-6-2757575 ext. 63320\\
Fax: +886-6-2766549 \\
Email: {\tt hshwchen@ieee.org}
\end{comment}

\date{\today}
\renewcommand{\baselinestretch}{1.2}
\thispagestyle{empty} \maketitle \thispagestyle{empty}
\newpage
\setcounter{page}{1}

\begin{abstract}
Nowadays, computation is playing an increasingly more important role in the future generation of computer and communication networks, as exemplified by the recent progress in software defined networking (SDN) for wired networks as well as cloud radio access networks (C-RAN) and mobile cloud computing (MCC) for wireless networks. This paper proposes a unified concept, i.e., computation diversity, to describe the impact and diverse forms of the computation resources on both wired and wireless communications. By linking the computation resources to the communication networks based on quality of service (QoS) requirements, we can show how computation resources influence the networks. Moreover, by analyzing the different functionalities of computation resources in SDN, C-RAN, and MCC, we can show diverse and flexible form that the computation resources present in different networks. The study of computation diversity can provide guidance to the future networks design, i.e., how to allocate the resources jointly between computation (e.g., CPU capacity) and communication (e.g., bandwidth), and thereby saving system energy and increase users' experiences.
\end{abstract}

%\newpage

\IEEEpeerreviewmaketitle

\vspace{0.3in}
\section{Introduction}
With ever increasing popularity of wired networks and mobile communications as well as the explosion of the applications such as multimedia gaming, high definition video, and virtual reality services, users are expecting to receive services with much better experience than before, which poses an increasingly high burden on the existing wired and wireless networks. New networking paradigms, such as software defined networking (SDN) \cite{McKeown} in wired networks and cloud radio access networks (C-RAN) \cite{China} as well as mobile cloud computing (MCC) \cite{5445167} in wireless networks, have been proposed recently and soon attracted a great deal of interest in academia and industry.

In those networks, the computation resources exit more in a form of centralized manner, rather than a traditionally distributed fashion, such as the computing resources in SDN controller, baseband unit (BBU), and mobile cloud. The key technique in the centralized computing is virtualization, which can make the computation resources dynamically configurable, scalable, sharable, and re-allocatable on demand. With the help of virtual machine and scaling techniques, the computation resources can present diverse and flexible capacity to meet different requirements of different wired and wireless communication networks. It has also been shown that computation resources can probably be provided by a general cloud pool, which is composed of the standard hardware of computing node (e.g., X86 sever), the standard storage, and the standard network switch, with the help of virtualization techniques in computing, storage, and networking. Thus, the computation can be treated as basic and fundamental resources to support not only wireless communications but also wired communications. However, computation resources are limited and would contribute to a large part of the total energy consumption. It has been shown that the energy overhead or the corresponding electricity costs are among the most important factors in the overall capital and operational expenditure in the network operators \cite{5493373}. Therefore, it is of great importance to study how computation resources present in different networks and how to allocate and utilize computation resources efficiently and effectively in those networks and communications.

In this paper, to explore the impact of the computation on different communication networks, we propose a unified concept, i.e., computation diversity, to describe the capacity and diverse forms of the computing resources in both wired and wireless networks. To study this, we analyse three emerging networking paradigms, i.e., SDN, C-RAN, and MCC, from network services, from network functions, and from users' experiences. Also we analyse the relationship between computation resources and communication resources in those networks.

The remainder of this article is organized as follows. Section II  introduces the concept of computation diversity. Sections III, IV, and V analyze the computation diversity in SDN, C-RAN, and MCC, respectively, followed by the suggestion of future networks design considering computation
diversity in Section VI. Finally, conclusion remarks are given in Section VII.

\vspace{0.3in}
\section{Computation Diversity}
Inspired by the concept of diversity in telecommunications, e.g., time diversity, frequency diversity, and space diversity, which refer to the abilities of signal detection by using two or more communication channels with different characteristics, we propose the computation diversity in this article, to describe the abilities and diverse forms of the computation resources in the communication networks. We link the impact of the computation resources and communication resources to the delay constraint or quality of service (QoS) of the networks and corresponding communications. In particular, computation diversity can be explained in following two ways. First, computation diversity can describe the ability or effect of computation resources on the networks. In this case, we study how many computation resources are needed in different networks, in order to complete the communication process under certain QoS requirements. Second, computation diversity can describe diverse forms of computation resources in different communication networks. We illustrate this by studying different presenting forms of computation resources in different networks, i.e., wired SDN and wireless C-RAN and MCC.

To be more specific, we first study computational capacity in the SDN controller as the first form of computation diversity. In this case, the computation resource is in charge of the decision making for the SDN switches and the processing of the instructions from the SDN applications. We then study the second form of computation diversity in the wireless C-RAN. In this case, the computation resources in BBU conduct the baseband signal processing and other calculations to guarantee the QoS of wireless communications. Then, we study the third form of computation diversity in MCC, where computation resources in mobile cloud is responsible for the execution of the tasks offloaded by mobile users and then transmitting the calculation results back to users through the networks.

By studying computation diversity, we can better understand the relations between communications and computation, and enable us to design joint resource allocation algorithms between them, then achieving the goal of saving energy and increasing the efficiency of the whole networking systems and the experiences of users. We summarize the network configurations and key notations of SDN, C-RAN, and MCC in Table I.

\begin{table*}
\caption{\label{tab1}Network configurations.}
\centering
\subtable[SDN]{
\begin{tabular}{|m{2.67cm}<{}|l|l|l|l|l|}
\hline
Configuration & Notation \\
\hline
\hline
Computational capacity in the SDN controller & $f^S$  \\
\hline
QoS  & $\tau$   \\
\hline
Packets arriving rate via SDN switches  & $\lambda$  \\
\hline
Packets no matching rate in flow table of SDN switches   & $p$   \\
\hline
Instruction rate from SDN applications & $\mu$ \\
\hline
\end{tabular}}
\qquad
\subtable[C-RAN]{
\begin{tabular}{|m{2.7cm}<{}|l|l|l|l|l|}
\hline
Configuration & Notation \\
\hline
\hline
Computational capacity in C-RAN BBU & $f^B$  \\
\hline
QoS  & $\tau$   \\
\hline
Number of antennas  & $A$   \\
\hline
Number of resource blocks  & $R$  \\
\hline
Modulation bits  & $M$   \\
\hline
Coding rate  & $C$ \\
\hline
Bandwidth allocated to the user & $B$  \\
\hline
\end{tabular}}
\qquad
\subtable[MCC]{
\begin{tabular}{|m{2.2cm}<{}|l|l|l|l|l|}
\hline
Configuration & Notation\\
\hline
\hline
Computational capacity in mobile clone & $f^C$  \\
\hline
QoS  & $\tau$   \\
\hline
Number of CPU cycles  & $F$   \\
\hline
Number of transmission data  & $D$  \\
\hline
Wireless data rate  & $r$   \\
\hline
\end{tabular}}
\end{table*}

\vspace{0.3in}
\section{Computation Diversity in Software Defined Networking}
The first type of diverse forms of the computation resource is presented in wired SDN. SDN is comprised of the SDN controller and SDN switch, which can decouple the control plane and data plane of the networks. Because of this feature, SDN can make networks directly programmable, manageable, controllable, and adaptable. Also, network administrators or applications can implement their services or introduce new algorithms quickly and easily via SDN.

\begin{figure}[htbp]
\centering
\includegraphics[width=6in]{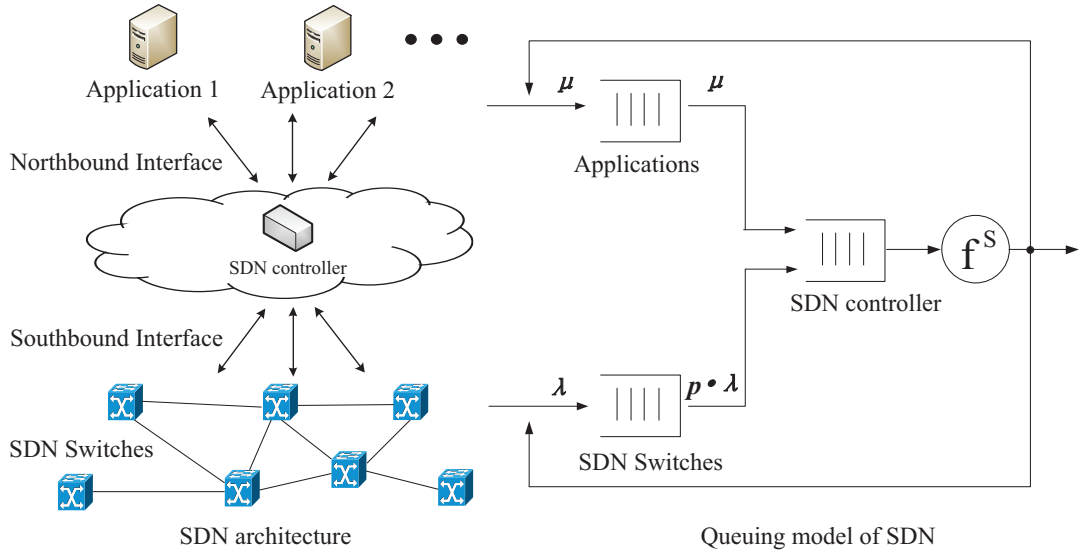}
\caption{SDN architecture and queuing model representation of SDN. } \label{sdnf-sdn}
\end{figure}

We show the SDN architecture in the left hand side of Fig. \ref{sdnf-sdn}, where the SDN controller conducts decision making for network flows by manipulating flow tables in the managed network element, i.e., SDN switch, where the traffic is sent from and forwarded to. In SDN, the controller can control and manage the data flow and switches via southbound interfaces. Moreover, if there is new entry of packets without finding routing in the flow table, these packets will be forwarded to the SDN controller via southbound interface for further processing. In SDN, a controller can provide the SDN applications with an abstract view of the networks, and therefore those applications can send new instructions via northbound to the controller to manage the network.

Therefore, one can see that the performance of the SDN depends largely on the processing capability of SDN controller, and thus if the SDN controller cannot process those requests fast enough, it will influence the responding time to the SDN switches and the applications, and then the quality of the whole networks. It may also cause packet loss or network failure in these situations. With the help of virtual machines and scaling techniques, the SDN controller can have better and diverse performance responding to the requests both from northbound and southbound. Therefore, one can see that the performance of the SDN controller depends mainly on the allocated computation resources, the instruction rate from the applications via northbound, and packets routing request rate from the SDN switch via southbound.

Inspired by the work \cite{6038457}, we abstract the SDN controller as a queuing system of the type M/M/1 to illustrate the computational capacity of SDN, as shown in right hand side of Fig. \ref{sdnf-sdn}. Specifically, we assume that the instructing rate of the SDN applications to the SDN controller from the northbound follows a Poisson process with its arrival rate $\mu$. Also, we assume that the data flow arriving rate in SDN switches follows a Poisson process with its arrival rate $\lambda$. There is possibility $p$ that the packets cannot find the forwarding path in the flow table and will be forwarded to controller for further processing. Therefore, the input rate from the southbound to the controller can be given as $p\cdot \lambda$. Moreover, assume that the service time at a controller follows an exponential distribution with its mean $\frac{1}{f^S}$. And then the processing delay in the controller can be given by $\frac{1}{f^S - p\cdot \lambda-\mu}$.

Next, we define the QoS of the networks as the delay requirement $\tau$, which means that the controller has to complete computation task no longer than time $\tau$ to avoid failure or packet loss in the networks. Different tasks may have different QoS requirements from time to time.
To statistically describe a task behaviour at runtime, a distribution such as exponential, normal, Weibull, and uniform can be assumed \cite{Jorge}.
In this section, we assume that $\tau$ follows a normal distribution $\mathcal{N}(\rho, \sigma^2)$ with the mean $\rho$ and the variance $\sigma^2$. Note that other distributions can also be applied but a similar performance can be observed. Using the popular $Q$-function \cite{JohnG}, we can define the failure rate or outage probability of the network as
\begin{equation}\label{1}
\begin{aligned}
P^{O} = 1-Q \Big( \frac{\frac{1}{f^S - p\cdot \lambda-\mu} -\rho}{\sigma^2}\Big).
\end{aligned}
\end{equation}

\begin{figure*}[htbp!]
\centering \subfigure[]{\includegraphics [width=3.2in]{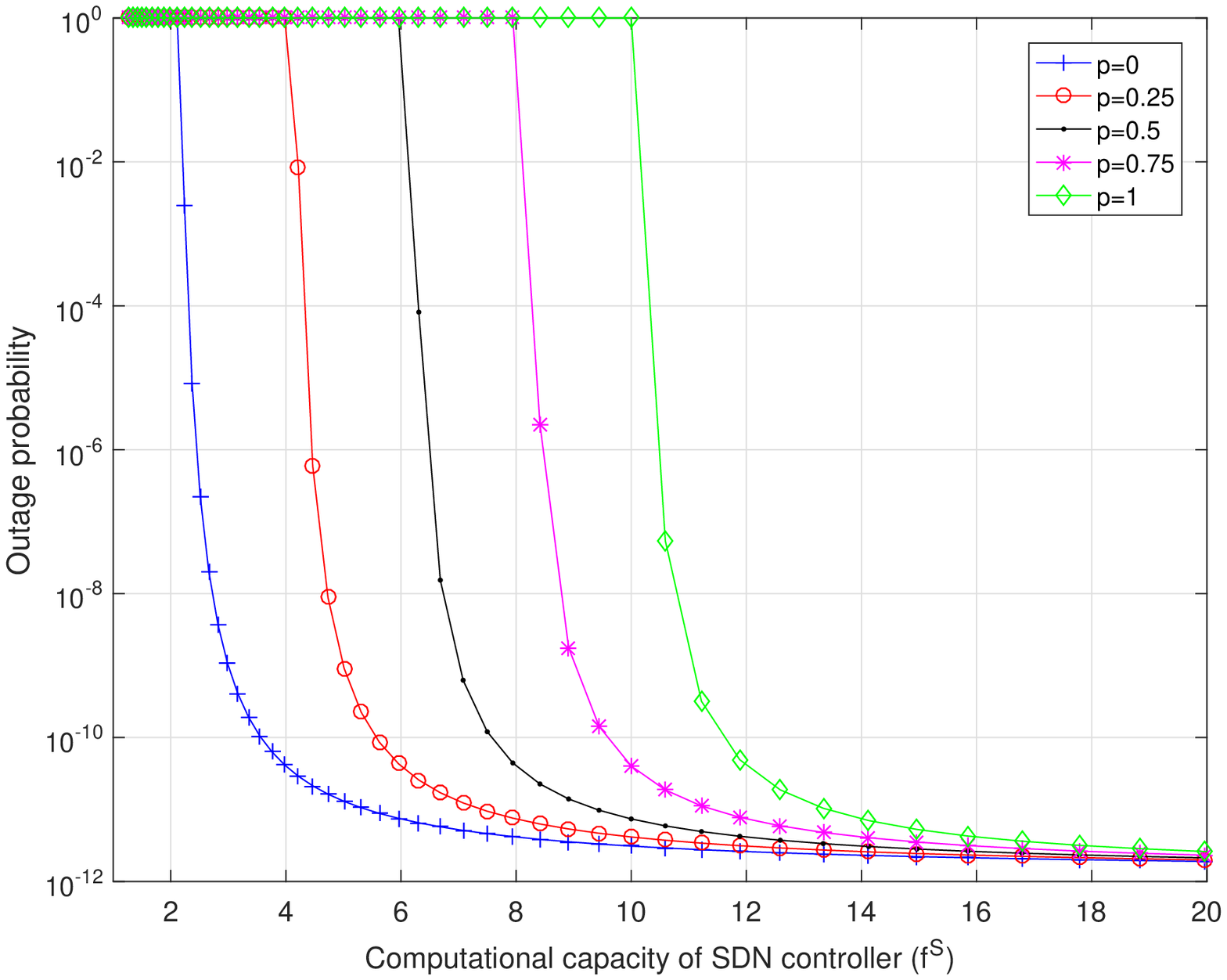}}
\subfigure[]{\includegraphics [width=3.2in]{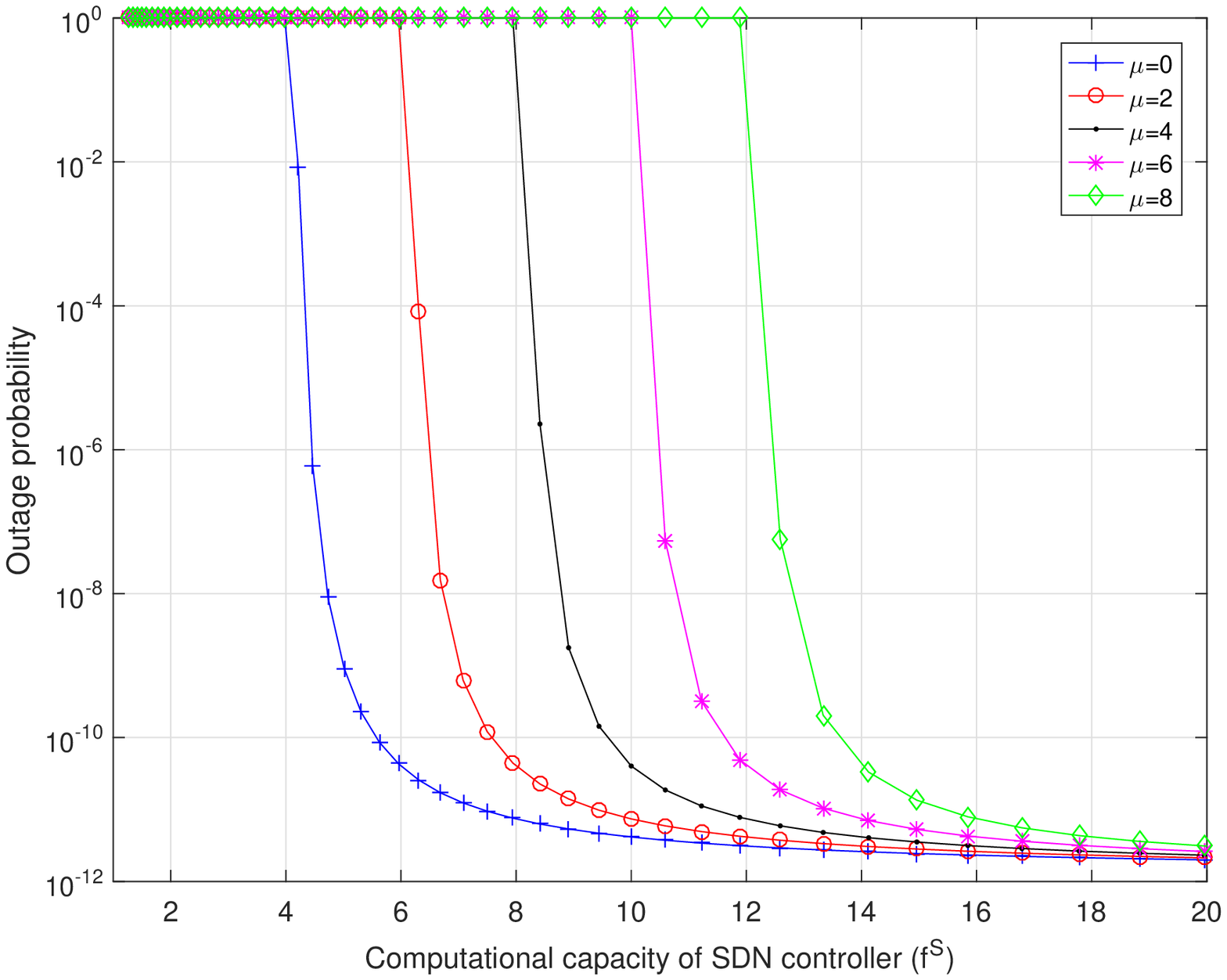}}
\caption{Outage probability versus computational capacity of SDN controller with (a) different values of $p$, (b) different instruction rates $\mu$. } \label{wired-wired2}
\end{figure*}

Using above equation, we show the simulation results in Fig. \ref{wired-wired2} to present the influence of the computation resources of the SDN controller to the networks. We set the arriving rate of SDN switch as $\lambda = 8$, and $\tau$ follows a normal distribution with the mean $\rho=7$ and the variance $\sigma^2=1$.

In Fig. \ref{wired-wired2} (a), we show the outage probability of the networks versus different computation capacities of the SDN controller with different $p$, and $\mu=2$ is set in this figure. One can see that with the increase of computational capacity of SDN controller, the outage probability is decreased, as expected. This is because the higher computational capacity the controller has, the lower probability the packets will be dropped. When the computational capacity $f^S$ is less than 2, the controller cannot complete the network task in required time even if there is no request from the switch (i.e., $p=0$), as the outage probability is 1 in this case. Also, one can see that with the increase of no matching rate in the SDN switch, the required capacity of SDN controller also increases, as expected.

In Fig. \ref{wired-wired2} (b), we show that the outage probability versus different computation capacities with different instruction rates $\mu$, and $p=0.5$ is set in this figure. One can see that with the increase of instruction rates from northbound, the outage probability increases under the same computational capacity of the SDN controller. However, with the increase of the computational capacity of SDN controller, the outage probability will decrease under the same other conditions, as expected.

To sum up, from Figs. \ref{wired-wired2} (a) and \ref{wired-wired2} (b), one can see that computational capacity of SDN controller affects the whole networks' performance, in a similar way as the network resource influences the networks. Thus, computation resource may be seen as the virtual network resource and should not be ignored in the future network designs.

\vspace{0.3in}
\section{Computation Diversity in Cloud Radio Access Network}
\begin{figure}[htbp]
\centering
\includegraphics[width=6in]{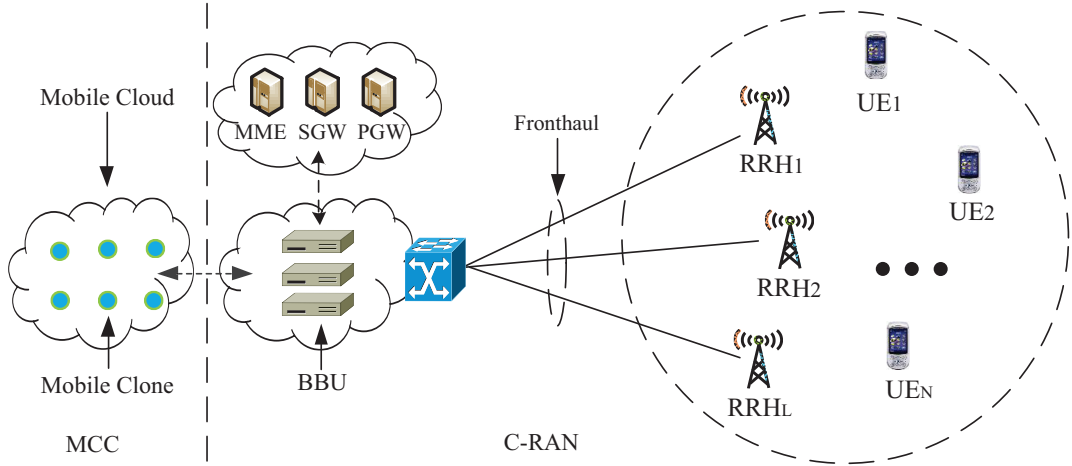}
\caption{C-RAN architecture (at the right hand side of the dashed line) and integrating C-RAN with MCC (at the left hand side of the dashed line). } \label{cran-cranbbu}
\end{figure}

In next-generation wireless access networks, a great emphasis is being placed on developing new networking technologies, such as small cells technology in 5G. However, small cell may not be able to handle the network processing and baseband signal processing tasks, such as precoding matrix calculation, channel state information estimation, fast Fourier transform (FFT), and forward error correction (FEC) in the required time. This is due to the possibility of lack of computation resources in small cells when a large amount of user equipments (UEs) appear in the same cell at the same time. To deal with this issue, another new networking architecture is proposed, i.e., C-RAN, which divides the traditional base station into remote radio heads (RRHs) and the BBU pool \cite{China}, as shown in the right hand side of the dashed line of Fig. \ref{cran-cranbbu}. In C-RAN, multiple BBU can be put together and realized by numerous software defined virtual machines in a cloud based data center.
RRH, which is similar to the SDN switch, can act as a soft relay to forward the received signals from BBU in the RF frequency band to UEs. Due to the feature of centralized signal processing in C-RAN, computation resources can be allocated very easily to BBUs to avoid insufficient provision of computation resources, and then prevent errors or packet loss to the large extent in wireless communications.

L. Guangjie, \emph{et al.} \cite{6477581} have provided an architecture of general processing processor (GPP) based BBU pool and studied the relationship between computation resources in BBU and the number of served UEs. Moreover, \cite{NavidNikaein} has shown that in order to meet the QoS of the transmission, the minimal CPU computational capacity has to be satisfied. Inspired by the above references, we would like to go a step further, to study how the computation resources in BBU affect the performance of the wireless communications, i.e., transmission data rate.

We define $f^B$ (Giga Operations Per Second, GOPS) as the computation resources required to serve the UE. From \cite{6214289, 7247587}, under certain QoS, one can see that $f^B$ can be expressed as $\frac{R}{10} (3 A+A^2+\frac{M C A}{3})$, where $R$ is the number of physical resource blocks allocating to the UE, $A$ is the number of used antennas, $M$ is the modulation bits, and $C$ is the code rate. Also, we can have the data rate $r$ as $\kappa^B  R M C$ in an OFDMA system with $\kappa^B$ setting to $1.68 \times 10^5$, if we assume 10\% overhead is used for reference signal, sync signals, PDCCH, etc. Also, assume the bandwidth allocating to the UE as $B$. Then, the number of physical resource blocks $R$ can be expressed as $\frac{B}{\kappa^A }$ with $\kappa^A$ setting to $2 \times 10^5$. Therefore, the relationship between the data rate and the computational capacity allocating to the UE can be given by
\begin{equation}\label{2}
\begin{aligned}
r=3 \;\kappa^B \bigg( 10\frac{f^B}{A}-3 \frac{B}{\kappa^A } -A \frac{B}{\kappa^A } \bigg).
\end{aligned}
\end{equation}

\begin{figure*}[htbp!]
\centering \subfigure[]{\includegraphics [width=3.2in]{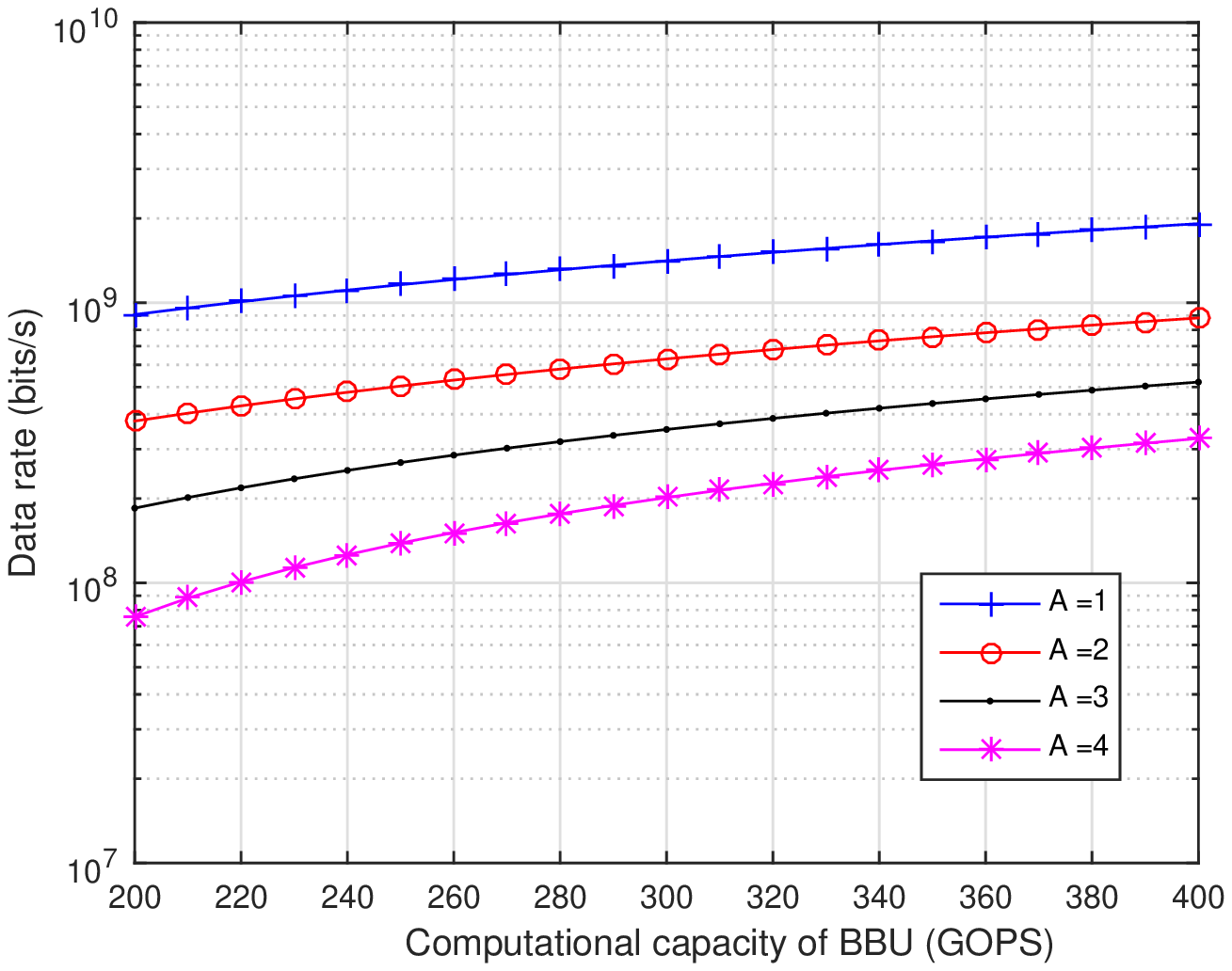}}
\subfigure[]{\includegraphics [width=3.2in]{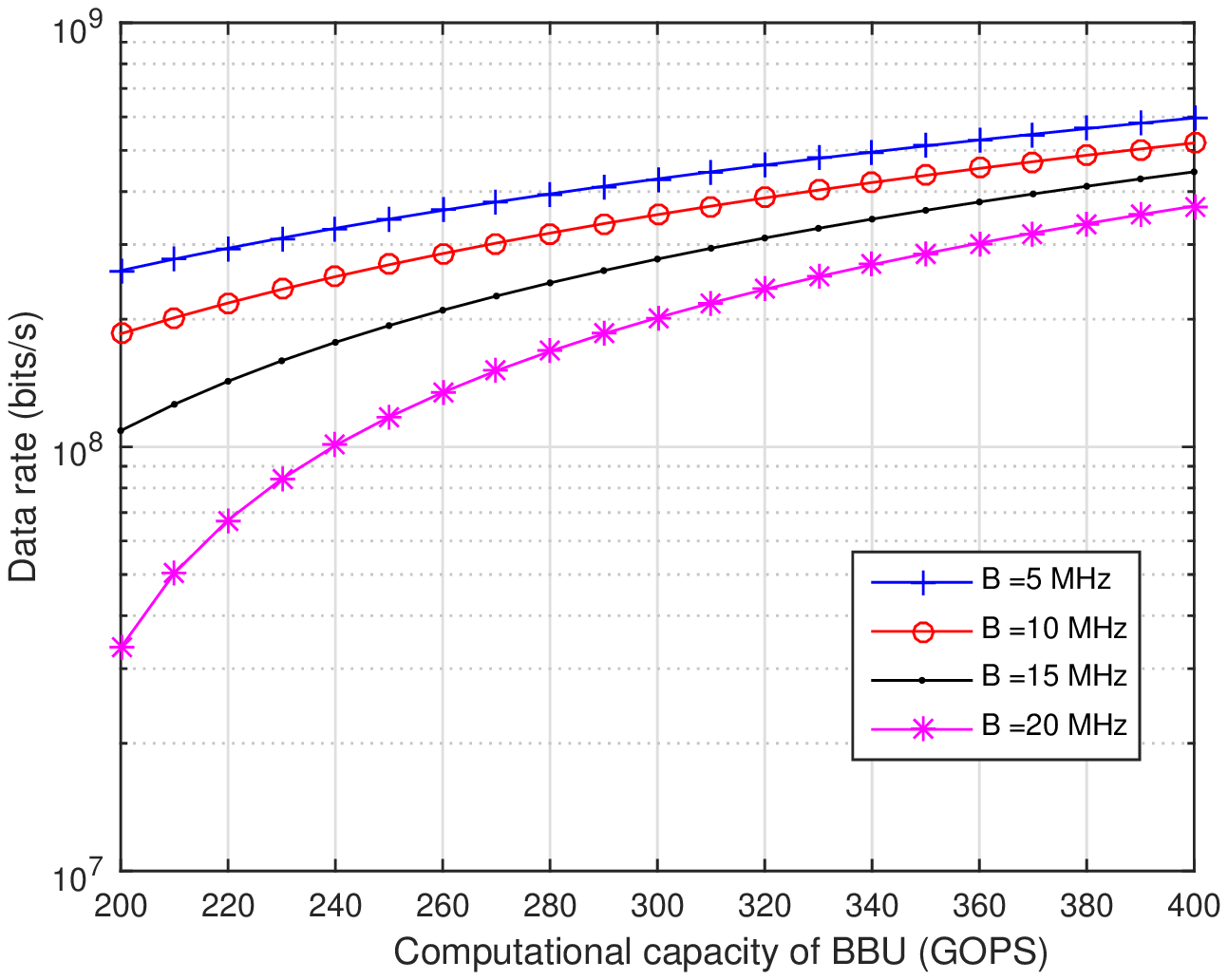}}
\caption{Date rate versus computational capacity of C-RAN BBU in the following two situations: (a) with different numbers of antennas with $B$=10 MHz bandwidth; (b) with different bandwidths with $A$=2 antennas.} \label{rifiAi-rifiBi}
\end{figure*}

In Fig. \ref{rifiAi-rifiBi}, we show the relationship between the wireless communication data rate and computational capacity of C-RAN. We set $B$=10 MHz in Fig. \ref{rifiAi-rifiBi} (a) and $A$=2 antennas in Fig. \ref{rifiAi-rifiBi} (b). In Fig. \ref{rifiAi-rifiBi} (a), one can see that with the increase of the computational capacity, the data rate also increases, as expected. Also, under the same data rate, with the increase of the number of the antennas, the computational capacity increases as well. This is because more antennas are employed, more data is needed to be processed, and then a higher computational capacity should be used. In Fig. \ref{rifiAi-rifiBi} (b), similarly, one can see that a higher computational capacity leads to a higher data rate. Also, under the same number of the antennas and data rate, with the increase of the bandwidth, a higher computational capacity is required. This is because more bandwidth means that more physical resource blocks are used, and then a higher computational capacity is employed to process them.

Overall, we show in this section that the allocated computational capacity in BBU influences the performance of wireless communications. Thus, we should not only consider the impact of wireless resources (e.g., bandwidth, etc.), but also consider the influence of computational resources in wireless systems. In other words, computational resources in wireless systems may be seen as the virtual wireless network resources, and should not be neglected in the future wireless system designs.

\vspace{0.3in}
\section{Computation Diversity in Mobile Cloud Computing}
The third form of the computation resources is observed from MCC. MCC is inspired by integrating the popular cloud computing into mobile environment, which enables that UEs with intensive computing demands but limited computation resources and batteries can offload their tasks to the powerful platforms in the cloud. In MCC, the mobile cloud is in charge of the task executing, and the wireless networks are responsible of receiving and transmitting the data from and to the UEs. The difference between MCC and normal cloud computing is that MCC has also to consider the wireless channel situation while meeting users' QoS requirements. In other words, both the wireless resources in the network and computation resources in the cloud affect UE's experience. For instance, if some tasks from the UE are needed to be completed in certain amount of time (i.e., QoS) in order to satisfy UE's experience, therefore, the UE which is allocated less communication resources by MCC may need more computation resources to calculate the task faster in order to meet the overall time constraints (task execution time plus data transmission time).

Practically, several cloud offloading platforms have been presented, such as Cloudlet \cite{5280678} and ThinkAir \cite{6195845}. Theoretically, \cite{5445167} has studied in different situations whether the offloading strategy to the cloud can save users' energy. However, in above works, they did not really consider how the computation resources affect the communications and the users' experiences. In MCC, UE only requires to receive the task results in certain time interval and therefore it is the system's responsibility to decide how many computation resources allocated to the task execution in the cloud and how many communication resources allocated to the data transmission in wireless networks, to the interest of the mobile operators, such as for the sake of minimization of energy, cost, etc.

To better illustrate the relationship between communication resources and computation resources in MCC, we propose MCC integrating with C-RAN architecture \cite{7511044, 7393804}, by adding mobile cloud in the left hand side of the dashed line of Fig \ref{cran-cranbbu}. In this structure, we assume that each UE has a specific mobile clone in the cloud. The mobile clone can be implemented by the cloud-based virtual machine, which holds the same software stacks, such as operating system, middleware, applications, as a corresponding UE. If the UE wants to execute computational intensive task, it will send the corresponding data to the mobile clone, which executes those tasks on mobile user's behalf. After task execution, the mobile clone will transmit the computation results back to the mobile user via C-RAN.

Similar to \cite{5445167}, we assume that UE has the computational intensive tasks to be accomplished with the parameters as $(F, D)$, where $F$ describes the total number of the CPU cycles required to be executed, while $D$ denotes the whole size of the data interaction between user and cloud through C-RAN, including the task's input and output data and the calculation results, etc. We also define the computational capacity of the mobile clone as $f^C$. Therefore, we can get the total time spending to complete this task as $\frac{F}{f^C} + \frac{D}{r}$, where the first term is the task execution time and the second term denotes data transmission time. If we assume the QoS requirement of the task as $\tau$, then we can get the relation between the computation resources in mobile clone and in communication as follows.
\begin{equation}\label{3}
\begin{aligned}
f^C=\frac{3 F \kappa^B (A^2 B+3A B-10 f^B \kappa^A)}{3 \kappa^B \tau ( A^2 B+3 A B-10 f^B \kappa^A)+A \cdot D \kappa^A},
\end{aligned}
\end{equation}
where equation (\ref{2}) in the last section is applied here and the parameters are defined before.

\begin{figure*}[htbp!]
\centering \subfigure[]{\includegraphics [width=3.2in]{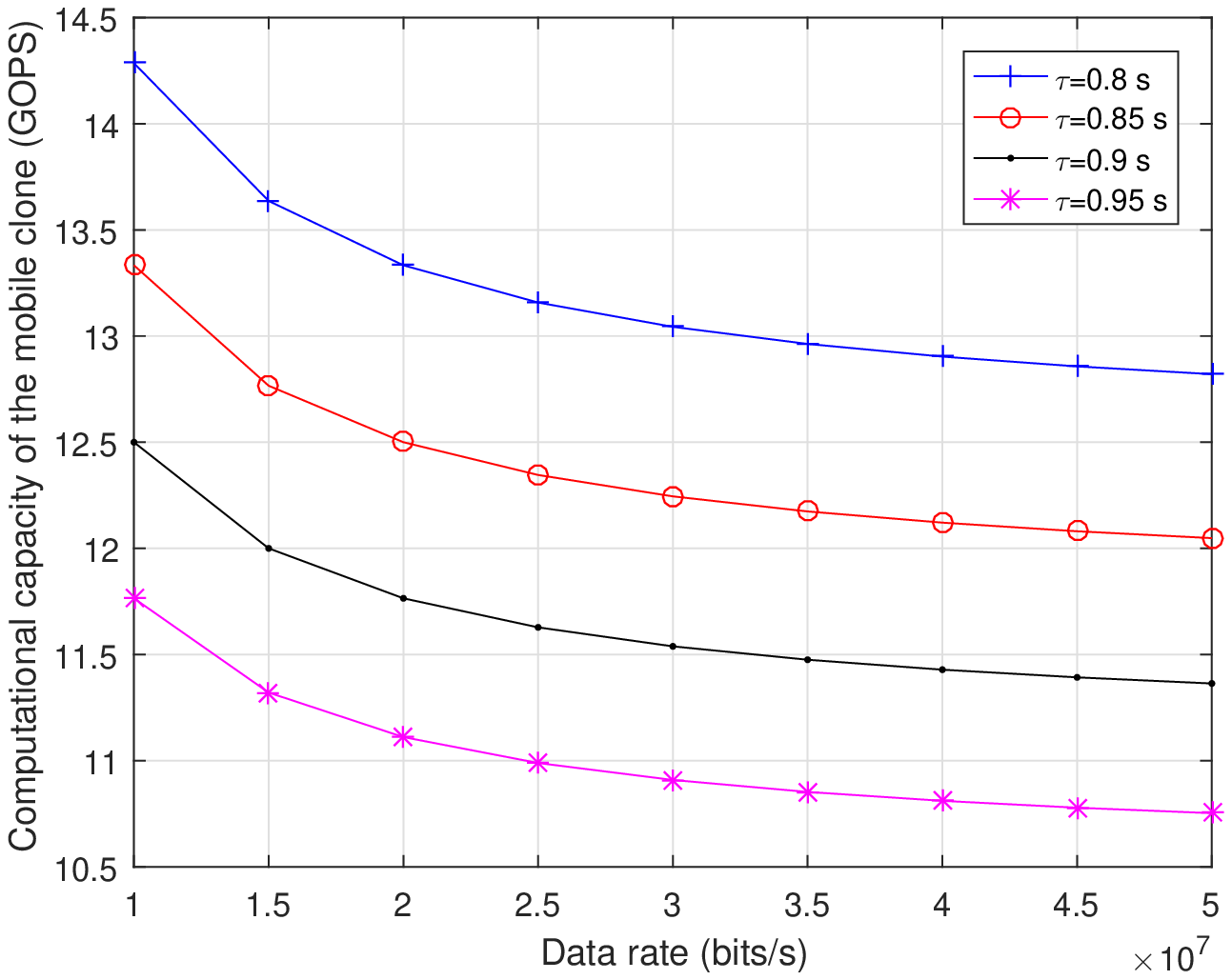}}
\subfigure[]{\includegraphics [width=3.2in]{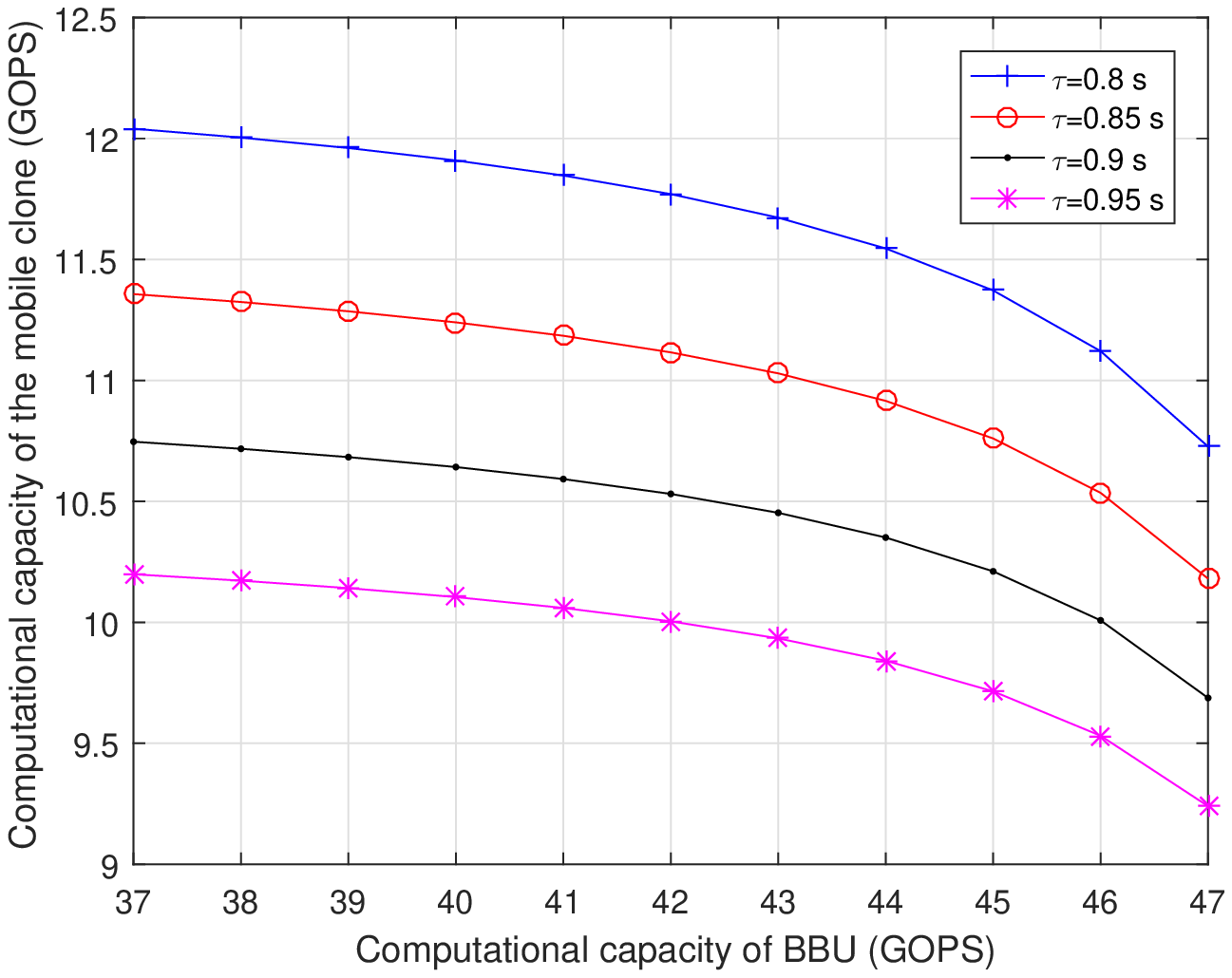}}
\caption{Computational capacity of mobile clone (a) versus date rate under $D=1$ Mb and $F= 10$ GOPS; (b) versus computational capacity of BBU under $B=10$ MHz, $A$=2, $D=1$ Mb and $F= 10$ GOPS.} \label{ficri-ficfib}
\end{figure*}

In Fig. \ref{ficri-ficfib} (a), we show the relationship between computational capacity of the mobile clone and wireless transmission data rate with $D=1$ Mb and $F=10$ GOPS under different QoS requirements. One can see that with the decrease of the time constraint, the computation resource required in mobile clone also increases. This is because under the same transmission data rate, less time constraints means that faster task execution is required, and then more computation resources in mobile clone are needed. Also, we can see with the increase of the transmission data rate, the computational capacity of the mobile clone can be reduced. This is because under the overall QoS requirement, a higher data rate means that more communication resources are allocated, and therefore less computation resources in mobile clone are needed.

In Fig. \ref{ficri-ficfib} (b), we show the relations between computational capacity of the mobile clone and the computational capacity of BBU with $D=1$ Mb and $F=10$ GOPS. The number of the antennas $A=2$ and the bandwidth $B=10$ MHz are set in this figure. One can see that with the increase of the computation resources in BBU, the computation resources in mobile clone can be reduced. Similarly, more computation resources in mobile clone lead to lower requirement of computation resources in BBU under the same QoS constraints. This is because if MCC allocates a higher computational capacity to mobile clone in task execution, then it can allocate less computation resources to BBU.

To summarize, in this section, we show the impact of computation resources on MCC. More specifically, we studied the relation between the service computation resources (in mobile clone) and communication computation resources (in BBU) under given QoS requirements. In this sense, computation resources in mobile clone can be treated as the virtual communication resources in MCC, as it also influences the network performance and user experience, in a similar way as the influence of communication resources to the network and to the users.

\vspace{0.3in}
\section{Possible Future Network Designs Considering Computation Diversity}

In future, networks may have diverse requirements and characteristics as well as different QoS demands for different applications. Thanks to the development of virtualization techniques, computational resources can be allocated more easily, diversely, and flexibly, by scaling up or scaling down the size of the virtual machine. Therefore, how to apply computation resources effectively and efficiently to the communication networks becomes a hot research topic. In future network designs, three possible directions are envisioned in this article, i.e., from the direction of network services, network functions, and users' experiences as follows.
\begin{enumerate}
\item
In the network services direction, SDN can be employed to decouple the network services from the underlying physical infrastructure. This feature enables applications or network administrators to manage network services conveniently through abstraction/virtualization of networks via SDN controller. By predicting and estimating the flows of networks and the instruction rate from the applications, operators are able to allocate computation resources properly to the networks. Also, SDN controller may be possibly implemented in virtual machine to increase its flexibility.
\item
In the network functions direction, by using the technique of network function virtualization (NFV), C-RAN BBU can be configured and tailored to realize different functions, according to the diverse requirements of the networks and the users. By implementing BBU in cloud-based virtual machine, computation resources can be allocated flexibly and conveniently in response to the change of the networks. Therefore, a proper allocation algorithm should be designed to efficiently assign computing resources to the networks, by analyzing the relations between computation resources and different network functions.
\item
The ultimate goal of the network design is to serve and satisfy the users. Therefore, from the users' experiences perspective, MCC can be possibly introduced in future networks architecture. In MCC, mobile clone can be seen as the virtualization of the users. By analyzing the relationship between QoS requirements of the tasks and computation along with communication, joint resource allocation can be developed and applied in the future network designs.
\end{enumerate}

\vspace{0.3in}
\section{Conclusions}

From above discussions about SDN, C-RAN, and MCC, we can see that computation is playing a very important role in recent emerging networking paradigms. By studying the diverse forms and different functionalities of the computation resources in different networks, (i.e. computation diversity), we can better design the future network architecture and allocate the right amount of computation resources to the networks. In this case, we can not only ensure the networks working properly and users having a better experience, but also save networking operators' resource and cost to a large extent.

\vspace{0.3in}
\section{Acknowledgements}
This work was supported in part by the U.K. EPSRC NIRVANA project (EP/L026031/1), EU Horizon 2020 iCIRRUS project (GA-644526), EU FP7 project CROWN (GA-2013-610524). Taiwan Ministry of Science and Technology project (104-2221-E-006-081-MY2), and the National Natural Science Foundation of China (61572389, 61572191).

\vspace{0.3in}
\bibliographystyle{ieeetran}
\bibliography{bare_jrnl}

% Generated by IEEEtran.bst, version: 1.14 (2015/08/26)
\begin{thebibliography}{10}
\providecommand{\url}[1]{#1}
\csname url@samestyle\endcsname
\providecommand{\newblock}{\relax}
\providecommand{\bibinfo}[2]{#2}
\providecommand{\BIBentrySTDinterwordspacing}{\spaceskip=0pt\relax}
\providecommand{\BIBentryALTinterwordstretchfactor}{4}
\providecommand{\BIBentryALTinterwordspacing}{\spaceskip=\fontdimen2\font plus
\BIBentryALTinterwordstretchfactor\fontdimen3\font minus
  \fontdimen4\font\relax}
\providecommand{\BIBforeignlanguage}[2]{{%
\expandafter\ifx\csname l@#1\endcsname\relax
\typeout{** WARNING: IEEEtran.bst: No hyphenation pattern has been}%
\typeout{** loaded for the language `#1'. Using the pattern for}%
\typeout{** the default language instead.}%
\else
\language=\csname l@#1\endcsname
\fi
#2}}
\providecommand{\BIBdecl}{\relax}
\BIBdecl

\bibitem{McKeown}
N.~McKeown and et~al, ``Openflow: Enabling innovation in campus networks,''
  \emph{SIGCOMM Comput. Commun. Rev.}, vol.~38, no.~2, pp. 69--74, 2008.

\bibitem{China}
C.~M.~R. Institute., ``{C-RAN} white paper: The road towards green ran.
  [online],'' \emph{(Jun. 2014)}, Available: http://labs.chinamobile. com/cran.

\bibitem{5445167}
K.~Kumar and Y.-H. Lu, ``Cloud computing for mobile users: Can offloading
  computation save energy?'' \emph{Computer}, vol.~43, no.~4, pp. 51--56, April
  2010.

\bibitem{5493373}
A.~Beloglazov and R.~Buyya, ``Energy efficient resource management in
  virtualized cloud data centers,'' in \emph{2010 10th IEEE/ACM International
  Conference on Cluster, Cloud and Grid Computing (CCGrid)}, May 2010, pp.
  826--831.

\bibitem{6038457}
M.~Jarschel, S.~Oechsner, D.~Schlosser, R.~Pries, S.~Goll, and P.~Tran-Gia,
  ``Modeling and performance evaluation of an openflow architecture,'' in
  \emph{2011 23rd International Teletraffic Congress (ITC)}, Sept 2011, pp.
  1--7.

\bibitem{Jorge}
J.~Cardoso, A.~Sheth, J.~Miller, J.~Arnold, and K.~Kochut, ``Quality of service
  for workflows and web service processes,'' \emph{Web Semantics: Science,
  Services and Agents on the World Wide Web}, vol.~1, no.~3, pp. 281 -- 308,
  2004.

\bibitem{JohnG}
J.~G. Proakis and M.~Salehi, \emph{Digital Communications}.\hskip 1em plus
  0.5em minus 0.4em\relax New York: McGraw-Hill, 2008.

\bibitem{6477581}
L.~Guangjie, Z.~Senjie, Y.~Xuebin, L.~Fanglan, N.~Tin-fook, Z.~Sunny, and
  K.~Chen, ``Architecture of {GPP} based, scalable, large-scale {C-RAN BBU}
  pool,'' in \emph{2012 IEEE Globecom Workshops (GC Wkshps)}, Dec 2012, pp.
  267--272.

\bibitem{NavidNikaein}
N.~Nikaein, ``Processing radio access network functions in the cloud: Critical
  issues and modeling,'' in \emph{Proceedings of the 6th International Workshop
  on Mobile Cloud Computing and Services}, 2015, pp. 36--43.

\bibitem{6214289}
C.~Desset, B.~Debaillie, V.~Giannini, A.~Fehske, G.~Auer, H.~Holtkamp,
  W.~Wajda, D.~Sabella, F.~Richter, M.~J. Gonzalez, H.~Klessig, I.~Gódor,
  M.~Olsson, M.~A. Imran, A.~Ambrosy, and O.~Blume, ``Flexible power modeling
  of {LTE} base stations,'' in \emph{2012 IEEE Wireless Communications and
  Networking Conference (WCNC)}, April 2012, pp. 2858--2862.

\bibitem{7247587}
T.~Werthmann, ``Approaches to adaptively reduce processing effort for {LTE}
  {C}loud-{RAN} systems,'' in \emph{2015 IEEE International Conference on
  Communication Workshop (ICCW)}, June 2015, pp. 2701--2707.

\bibitem{5280678}
M.~Satyanarayanan, P.~Bahl, R.~Caceres, and N.~Davies, ``The case for vm-based
  cloudlets in mobile computing,'' \emph{IEEE Pervasive Computing}, vol.~8,
  no.~4, pp. 14--23, Oct 2009.

\bibitem{6195845}
S.~Kosta, A.~Aucinas, P.~Hui, R.~Mortier, and X.~Zhang, ``Thinkair: Dynamic
  resource allocation and parallel execution in the cloud for mobile code
  offloading,'' in \emph{2012 IEEE Proceedings INFOCOM}, March 2012, pp.
  945--953.

\bibitem{7511044}
K.~Wang, K.~Yang, X.~Wang, and C.~S. Magurawalage, ``Cost-effective resource
  allocation in c-ran with mobile cloud,'' in \emph{2016 IEEE International
  Conference on Communications (ICC)}, May 2016, pp. 1--6.

\bibitem{7393804}
K.~Wang, K.~Yang, and C.~Magurawalage, ``Joint energy minimization and resource
  allocation in {C-RAN} with mobile cloud,'' \emph{IEEE Transactions on Cloud
  Computing}, vol.~PP, no.~99, pp. 1--1, 2016.

\end{thebibliography}

\end{document}